\documentclass[prl,twocolumn,superscriptaddress,
amsmath,amssymb,epsfig]{revtex4}
\usepackage{epsf}
\usepackage{amsmath}
\usepackage[english]{babel}
\usepackage{amssymb}
\usepackage[applemac]{inputenc}
\usepackage{graphicx}
\usepackage{syntonly}

\newcommand{\ket}[1]{ \left
|#1\right\rangle}
\begin{document}
\title{Non-Probabilistic Termination of Measurement-based Quantum Computation}
\author{Philippe Jorrand}
\email{philippe.jorrand@imag.fr}
\author{Simon Perdrix}
\email{simon.perdrix@imag.fr}
\affiliation{Leibniz Laboratory\\46 avenue Félix Viallet 38000 Grenoble, France}

\begin{abstract}
Nielsen \cite{NIE} introduced a model of quantum computation by measurement-based simulation of unitary computations. In this model, a consequence of the non-determinism of quantum measurement is the probabilistic termination of simulations. This means that the time when simulation terminates for a given computation is probabilistic, and this simulation may even never end.

We introduce (section 3) a measurement-based model with non probabilistic termination, which permits, unlike existing models, to predict the time of termination. This new scheme is a modification of Nielsen's. After an introduction to Nielsen's scheme (section 1), an analysis of different temporal organisations of elementary simulations within Nielsen's scheme (section 2) leads to the non probabilistic model. 
\end{abstract}

\date{\today}
\maketitle

\section{Nielsen's Scheme}
\subsection{The Goal}

The goal of Nielsen's scheme \cite{NIE} for measurement-based simulation of quantum computation is to prove the universality of quantum measurement, i.e. that any unitary operator $U$ can be simulated using only quantum measurements. 
\begin{description}
\item[Nielsen's Scheme:]
\end{description}
\begin{itemize}
\item The first stage consists in decomposing any unitary operation $U$ into members of a universal family of unitary operators. In this article the $\{One-qubit\ unitary\ transformation, CNot\}$ universal family \cite{NC} is chosen first.

\item The second stage consists in simulating each member of the universal family using only quantum measurements. These elementary simulations are performed using generalised teleportation.
\end{itemize}

\subsection{Generalised Teleportation}

\subsubsection{From teleportation without unitary operation\ldots}
The following network can be considered as a quantum teleportation scheme:

\begin{center}
\includegraphics[width=0.45\textwidth]{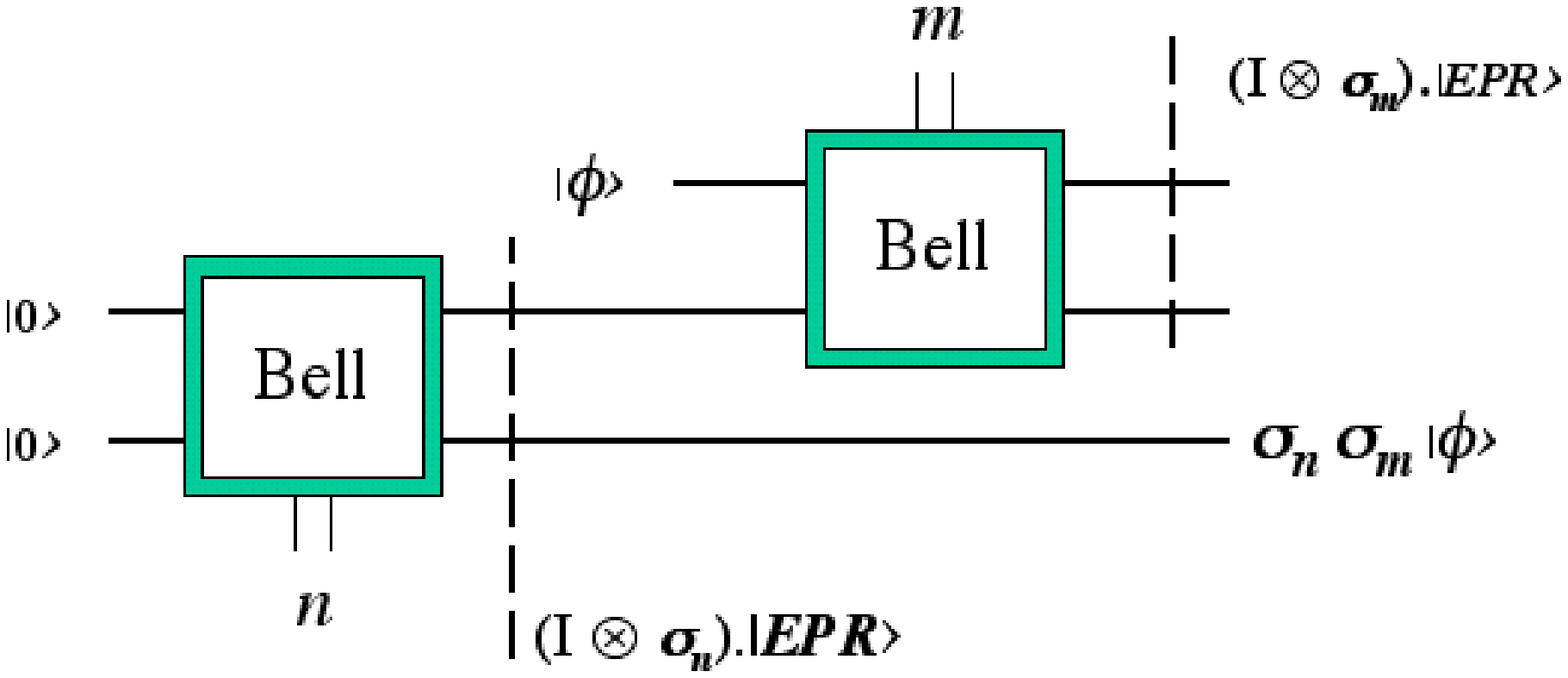}
\emph{\small figure 1}
\end{center}

Bell measurements are two-qubit measurements in Bell's basis (i.e. the basis of $C^4$ composed of the four Bell states). Each Bell state can be expressed as $(Id\otimes \sigma_n)\ket{EPR}$, for $n\in \{0..3\} $, where $\ket{EPR}=\frac{\ket{00}+\ket{11}}{\sqrt{2}}$ and $\sigma_n$ is a Pauli operator.

The first Bell measurement creates an entangled Bell state. The action of the second Bell measurement can be viewed as equivalent to the application of $H$ then $CNot$ (see \emph{figure 3}) followed by standard measurement, as in the usual presentation of quantum teleportation \cite{BB}. The application of the second Bell measurement teleports an unknown state $\ket{\phi}$ from Alice to Bob, up to a $\sigma_i=\sigma_n\sigma_m$ Pauli operator which depends on the classical results of the two successive measurements.  
\subsubsection{\ldots to the simulation of one-qubit unitary operations.}

Teleportation is nothing but a simulation of the \emph{identity} transformation because the final state is the same as the initial state. Nielsen showed that any one-qubit unitary transformation can be simulated, by adjusting the measurement basis.

\begin{center}
\includegraphics[width=0.45\textwidth]{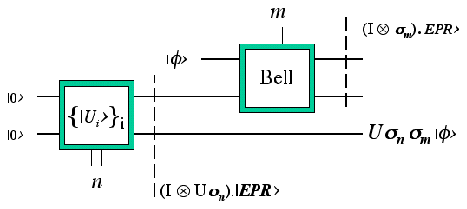}
\emph{\small figure 2}
\end{center}

For a given one-qubit unitary operator $U$, the set of states $\{\ket{U_i}=(I\otimes U\sigma_i)\ket{EPR}\}_{i=0..3}$ is a basis of the two-qubit Hilbert space. Thus, if the first Bell measurement of the teleportation scheme is replaced by a measurement in a basis $\{\ket{U_i}\}_{i=0..3}$ (\emph{figure 3}), then the application of this scheme to a state $\ket{\phi}$ generates a state $U\sigma_n\sigma_m\ket{\phi}$, where $n$  and $m$ are classical results of the measurements. 

\subsubsection{Corrective Stage}

Using the simulation presented above, a corrective stage is needed, due to the non-determinism of measurement. Whereas the desired result is $U\ket{\phi}$, the simulation produces $U\sigma_n\sigma_m\ket{\phi}$. If $m=n$, then the simulation is terminated, otherwise a corrective stage needs to be performed.

Nielsen proposes a corrective stage organised as follows:

\begin{description}
\item[Simulation Step:] In order to simulate the application of $U$ to a state $\ket{\phi}$ , the generalised scheme of teleportation is applied with $U_0 = U$ and $\ket{\psi_0} =\ket{\phi}$. The quantum output after this first simulation is $U_0\sigma_{m_0}\sigma_{n_0}\ket{\psi_0}$. 

\item[Correction Step:] If $m_0=n_0$ then the corrective phase is terminated. Otherwise the scheme is applied again with a simulation of $U_1 =U_0\sigma_{m_0}\sigma_{n_0}U_0^{\dag}$ on the state produced by the above simulation, which is $\ket{\psi_1} =U_0\sigma_{m_0}\sigma_{n_0}\ket{\psi_0}$. And so on.
\end{description}

At any step, the probability of success, which does not depend on $U$ and $\ket{\phi}$, is $1/4$. Therefore, the probability that the number $x$ of calls to the scheme of generalised teleportation be greater than $k$, is $P(x>k)=(3/4)^k\xrightarrow[k\rightarrow \infty]{} 0$. This result shows that the probability that a simulation never ends is null, but it does not permit to predict a date after which a simulation is terminated with probability one: this model is a \emph{probabilistic terminating model}.

\subsubsection{Extension to a simulation of CNot}

The above scheme simulates any one-qubit unitary transformation. Nielsen extended this scheme to a simulation of any two-qubit unitary transformation \cite{NIE}. This extension needs four-qubit measurements.
Then Leung \cite{LEU1,LEU2} showed that a simulation of $CNot$ can be obtained using only two-qubit measurements. Due to the universality of $\{$\emph{One-qubit unitary transformation},$ CNot\}$, and since simulation of any one-qubit unitary operator needs only two-qubit measurements, Leung concluded to the universality of two-qubit measurements.

\section{Postponement of Corrective Stage}
\subsection{Motivations}

Using Nielsen's scheme, each simulation step is followed by an average of four steps of correction. That is why one can wonder if such resource costly corrections may be better organized. An alternative scheme to Nielsen's consists in postponing all corrective steps into a global corrective stage. The body of the simulation stage is thus composed of simulation steps only.

For instance, consider the simulation of a unitary transformation $U=(H\otimes I)CNot$ on a state $\ket{\phi}$. After the first step of simulation a state $CNot(\sigma_n\otimes\sigma_m)\ket{\phi}$ is obtained, where $n$ and $m$ depend on the results of the measurements. Whereas Nielsen's scheme goes on by correcting this state until obtaining $CNot\ket{\phi}$, an alternative scheme consists in the simulation of $H$ without any preliminary corrective step. This simulation leads to a state $U_{simul}\ket{\psi}=(H\sigma_k\otimes I)CNot(\sigma_n\otimes\sigma_m)\ket{\phi}$, where $k$ depends on the results of the measurements performed during the simulation of $H$. The global corrective stage may be represented by a unitary operator $C_U=UU^{\dag}_{simul}$,  thus $C_UU_{simul}\ket{\phi}=UU^{\dag}_{simul}U_{simul}\ket{\phi}=U\ket{\phi}$.

In the general case, the corrective operator $C_U$ is obtained using the definition $C_U=UU^{\dag}_{simul}$, where $U$ is the simulated unitary operator, and $U_{simul}$ is the operator actually simulated during the simulation stage. Knowing $U$, $U_{simul}$ is entirely determined by the results of the measurements performed during the simulation stage.

\subsection{Discussion around Benefit of Postponement}

The existence of a corrective operator $C_U$ for any unitary operator $U$ allows to run the whole simulation stage without taking care of classical results of measurements. In this way the simulation stage can be considered as \emph{unconditional}.

To point out the benefit of this alternative scheme, one has to compare, in terms of number of measurements performed, the simulation of $U$ using, on one hand the original Nielsen's scheme and on the other hand the postponement of the corrective stage, where the simulation of $C_U$ still has to obey Nielsen's scheme. Intuitively, the benefit of the postponement comes from the interactions between some corrective factors called for by the simulation stage.
The next section is dedicated to the characterisation of the gain due to the postponement strategy.

\section{Toward a Non-Probabilistic Model}
\subsection{An Attemp of Error Characterization}

To quantify the hypothetical gain due to the postponement  of the corrective stage, a universal approximation family $\{H, T, CNot\}$ is considered \cite{NC}: 
\begin{center}
$$H=\frac{1}{\sqrt{2}}\left( \begin{array}{cc}
1&1 \\
1&-1 
\end{array}
\right), \ \ T=\left( \begin{array}{cc}
1&0 \\
0&e^{\frac{i\pi }{4}} 
\end{array}
\right)$$

$$CNot=\left( \begin{array}{cccc}
1&0&0&0 \\
0&1&0&0 \\
0&0&0&1 \\
0&0&1&0 
\end{array}
\right)$$

\emph{\small figure 3}

\end{center}

The computation of any $U$ can be decomposed into $l$ steps, where each step consists in an application of $H$, $T$ or $CNot$. Step by step, simulations of $H$, $T$ and $CNot$ introduce errors, which are accumulated according to the postponement strategy. Thus there exists, for any $k$, a corrective operator $C_{k}$, which represents the errors accumulated during the $k$ first steps of simulation. Notice that $C_{l}$ is nothing but $C_U$. The computation simulated by the first $k$ steps is suppsedto produced a state $\ket{\psi_k}$, but the state actually coming out of these $k$ simulation steps is $\ket{\eta_k}$. Thus, for any $k$, $\ket{\psi_k}=C_{k}\ket{\eta_k}$.

The special case where, for any $k$, the corrective operator $C_k$ is a Pauli operator is interesting to consider:
\begin{description}
\item[Property 1 - ] After $k$ steps of simulation, the corrective operator $C_{k}$ is a $n$-qubit Pauli operator:
$$C_{k}=\alpha ( \bigotimes_{j=1..n} \sigma_{d_j})$$
where $\alpha$ is a global phase and $d_j\in\{0..3 \}$
\end{description}

\begin{description}
\item[Lemma 1 - ] After $k$ steps of simulation of $H$ or $CNot$, the corrective operator $C_{k}$ is a $n$-qubit Pauli operator.
\end{description}

{\bf Proof:} By induction, using the stabilization of the Pauli group (composed of the Pauli operators) by $CNot$ and $H$.\hfill $\Box$

Unfortunately, using the elementary simulation proposed by Nielsen, the result of a simulation of $T$ on a state $\ket{\psi}$ may be $T\sigma_1\ket{\psi}=\frac{1}{\sqrt{2}}(\sigma_1+\sigma_2)T\ket{\psi}$. Since $ \frac{1}{\sqrt{2}}(\sigma_1+\sigma_2)$ is not a Pauli operator, the elementary simulations introduced by Nielsen do not verify \emph{Property 1}.

In order to have this property satisfied also by $T$, an alternative elementary simulation of $T$ must be found.

\subsection{An Alternative Elementary Simulation of $T$}

Thanks to \emph{Lemma 1}, satisfying \emph{Property 1} (i.e. the corrective operator is a Pauli operator) is reduceable to finding an adapted simulation for $T$. A simulation of $T$ is \emph{adapted} if any state $C_k\ket{\phi}$ is transformed into $C_{k+1}T\ket{\phi}$, where $C_k$ and $C_{k+1}$ are Pauli operators. The transformation $T$ is a one-qubit unitary operator, so the previous condition can be rewritten as follow: a simulation of $T$ is \emph{adapted} if any state $\sigma_p\ket{\phi}$ is transformed into $C_TT\ket{\phi}$, where $C_T$ is a one-qubit Pauli operator.

The alternative elementary simulation of $T$ is based on \emph{projective measurements}, e.g. a $\sigma_3\otimes\sigma_3$-measurement. If the $4\times4$ Pauli matrix $\sigma_3\otimes\sigma_3$ is considered as a two-qubit observable, then a $\sigma_3\otimes\sigma_3$-measurement is a projective measurement, with eigenspaces $\{\ket{00},\ket{11}\}$ and $\{\ket{01},\ket{10}\}$.
A Bell measurement can be decomposed into two successive projective measurements: $\sigma_3\otimes\sigma_3$ then $\sigma_1\otimes\sigma_1$.
The measurement $\sigma_3\otimes\sigma_3$ then $\sigma_2\otimes\sigma_1$ is  close to a Bell measurement, since the eigenvectors of this composed measurement are $\frac{(\ket{00}+i\ket{11})}{\sqrt{2}}$, $\frac{(\ket{01}+i\ket{10})}{\sqrt{2}}$, $\frac{(\ket{00}-i\ket{11})}{\sqrt{2}}$, $\frac{(\ket{01}-i\ket{10})}{\sqrt{2}}$.

The general structure of the simulation of $T$ proposed by Nielsen is conserved, only the Bell measurement is replaced by a two-qubit operator $M$. 

\begin{center}
\includegraphics[width=0.45\textwidth]{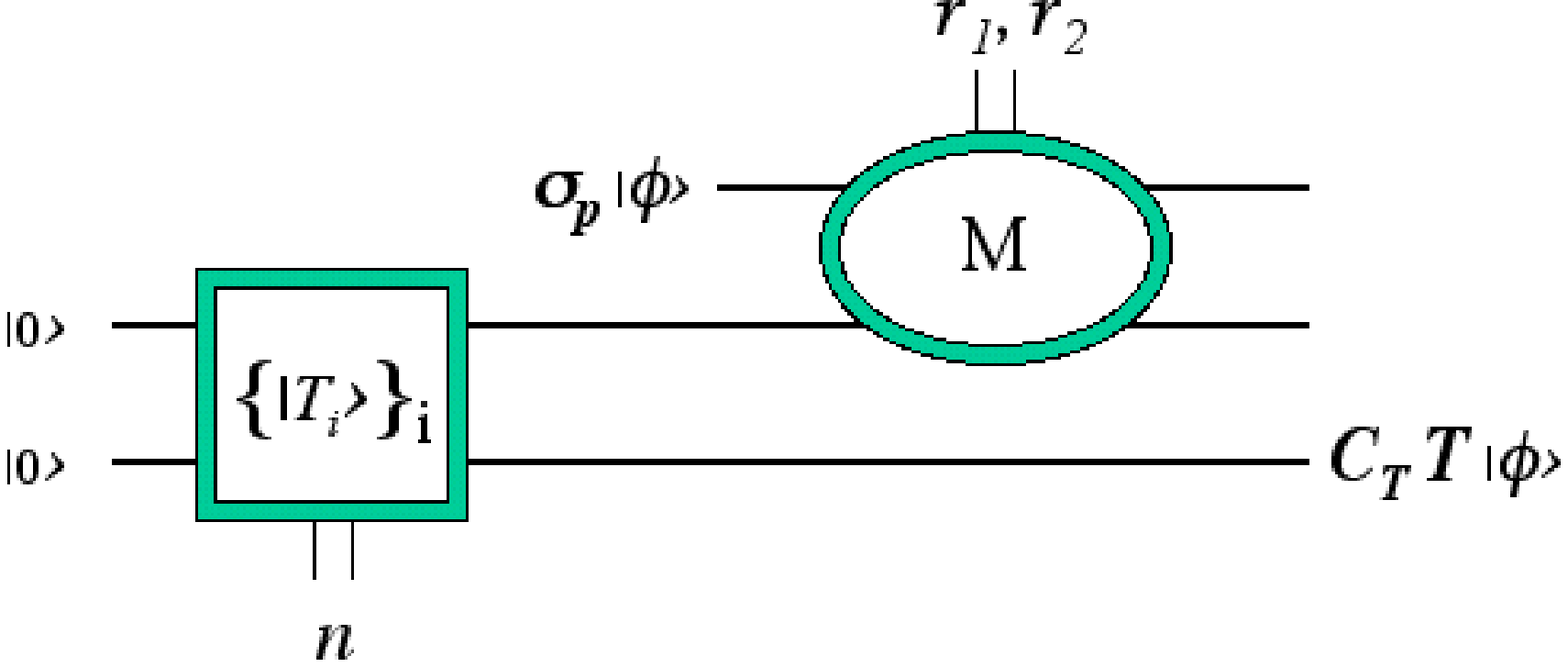}
\emph{figure 4}
\end{center}

Operator $M$ is composed of two successive projective measurements. The first projective measurement, $M_1$, depends on the corrective operator accumulated before the simulation (i.e. $\sigma_p$), and $n$, which is the result of the first measurement. The second projective measurement, $M_2$, depends on $\sigma_p$, $n$ and also on the result $r_1$ of the first projective measurement $M_1$. The full description of $M$ is given in \emph{table 1}.

For instance, the third line should be read as follows: if $\sigma_p=\sigma_0$ and $n=2$, then $M_1$ is a $-\sigma_3\otimes\sigma_3$-measurement. If the classical result $r_1$ of $M_1$ is $1$, then $M_2$ is a $\sigma_1\otimes\sigma_1$-measurement, otherwise (i.e. if $r_1=-1$) $M_2$ is a $\sigma_2\otimes\sigma_1$-measurement.

\begin{description}
\item[Theorem 1 - ] 
With the alternative elementary simulation of $T$, thecorrective operator $C_T$ is a Pauli operator.
\end{description}

{\bf Proof:} A calculative proof permits to show that:

- if $r_1=r_2=1$, then $C_T=\sigma_0$.

 - if $r_1=-1$ and $r_2=1$, then $C_T =\sigma_1$.

 - if $r_1=r_2=-1$, then $C_T =\sigma_2$.

 - if $r_1=1$ and $r_2=-1$, then $C_T=\sigma_3$.

$\hfill\Box$

{\tiny
\begin{center}
\begin{tabular}{|c|c|c|c|c|}
\hline

$\ \sigma_p\ $
&$\ n\ $
&$1^{st}$
&\multicolumn{2}{|c|}{$2^{nd}\ mesurement$} \\
&
&$measurement$
&$r_1=1$
&$r_1=-1$\\
\hline
$\sigma_0$&$0$&$\ \sigma_3\otimes\sigma_3$&$\sigma_1\otimes\sigma_1$&$\sigma_2\otimes\sigma_1$\\
\hline
$\sigma_0$&$1$&$-\sigma_3\otimes\sigma_3$&$\sigma_1\otimes\sigma_1$&$-\sigma_2\otimes\sigma_1$\\
\hline
$\sigma_0$&$2$&$-\sigma_3\otimes\sigma_3$&$\sigma_1\otimes\sigma_1$&$\sigma_2\otimes\sigma_1$\\
\hline
$\sigma_0$&$3$&$\ \sigma_3\otimes\sigma_3$&$-\sigma_1\otimes\sigma_1$&$\sigma_2\otimes\sigma_1$\\
\hline
$\sigma_1$&$0$&$-\sigma_3\otimes\sigma_3$&$\sigma_1\otimes\sigma_1$&$\sigma_2\otimes\sigma_1$\\
\hline
$\sigma_1$&$1$&$\ \sigma_3\otimes\sigma_3$&$\sigma_1\otimes \sigma_1$&$\sigma_2\otimes \sigma_1$\\
\hline
$\sigma_1$&$2$&$\ \sigma_3\otimes\sigma_3$&$-\sigma_1\otimes\sigma_1$&$-\sigma_2\otimes\sigma_1$\\
\hline
$\sigma_1$&$3$&$-\sigma_3\otimes\sigma_3$&$-\sigma_1\otimes\sigma_1$&$-\sigma_2\otimes\sigma_1$\\
\hline
$\sigma_2$&$0$&$-\sigma_3\otimes\sigma_3$&$-\sigma_1\otimes\sigma_1$&$-\sigma_2\otimes\sigma_1$\\
\hline
$\sigma_2$&$1$&$\ \sigma_3\otimes\sigma_3$&$-\sigma_1\otimes\sigma_1$&$-\sigma_2\otimes\sigma_1$\\
\hline
$\sigma_2$&$2$&$\ \sigma_3\otimes\sigma_3$&$\sigma_1\otimes\sigma_1$&$\sigma_2\otimes\sigma_1$\\
\hline
$\sigma_2$&$3$&$-\sigma_3\otimes\sigma_3$&$\sigma_1\otimes\sigma_1$&$\sigma_2\otimes\sigma_1$\\
\hline
$\sigma_3$&$0$&$\ \sigma_3\otimes\sigma_3$&$-\sigma_1\otimes\sigma_1$&$\sigma_2\otimes\sigma_1$\\
\hline
$\sigma_3$&$1$&$-\sigma_3\otimes\sigma_3$&$-\sigma_1\otimes\sigma_1$&$\sigma_2\otimes\sigma_1$\\
\hline
$\sigma_3$&$2$&$-\sigma_3\otimes\sigma_3$&$-\sigma_1\otimes\sigma_1$&$-\sigma_2\otimes\sigma_1$\\
\hline
$\sigma_3$&$3$&$\ \sigma_3\otimes\sigma_3$&$\sigma_1\otimes\sigma_1$&$-\sigma_2\otimes\sigma_1$\\
\hline
\end{tabular}

{\small \emph{table 1}}

\end{center}}

\subsection{Non-Probabilistic Model}

Using this new model composed of simulations of $H$ and $CNot$ proposed by Nielsen and the alternative elementary simulation of $T$, the simulation of any $n$-qubit unitary transformation $U$ on a state $\ket{\psi}$ produces a state $C_UU\ket{\psi}$, where $C_U$ is now a Pauli operator. Corrections may now be performed according to two strategies.

The first strategy consists in a simulation of $C_U$ using Nielsen's scheme, independently on each qubit. The cost of correction with this strategy is $O(n)$, whereas the cost of correction with Nielsen's scheme is $O(l)$, where $l$ is the number of elementary operators necessary for the simulation of $U$, in general $l\gg n$.

The second strategy relies on a distinction between two contexts:

- The context of \emph{Quantum Computation}, where quantum results are measured in the computational basis. Pauli operators stabilize the computational basis $\{\ket{0},\ket{1}\}$, so the corrective operator $C_U$ does not modify the measurement basis, but only the interpretation of the measurement. For instance, measurements of  $\ket{\phi}$, and $\sigma_1\ket{\phi}$ are equivalent, up to an inversion of the role of classical results $0$ and $1$.

- In all other cases, the corrective operator may be, for instance, classically transmitted (using a $n$-tuple of integers between 0 and 3) to the receiver of the quantum result. 

When this strategy is chosen, particularly in a context of quantum computation, the corrective stage disappears entirely. The model composed of simulations of $H$ and $CNot$ proposed by Nielsen and the alternative simulation of $T$, associated with this strategy is a \emph{Non-Probabilistic Model}, due to the following theorem:

\begin{description}
\item[Theorem 2 - Non-Probabilistic Model] 

The simulation of any unitary transformation with the non-probabilistic model always terminates, and the date of termination can be predicted. 
\end{description}

\section{Conclusion}
Nielsen showed the universality of quantum measurement for quantum computation, then Leung proved this universality under the constraint that only two-qubit measurements are allowed. Finally we prove the universality of two-qubit measurements under the constraint that the end of the computation must be predicted. Thus, although quantum measurement is probabilistic, we introduce a model of quantum computation, based on measurements only, on which the number of performed steps, for a given simulation, is not probabilistic. Moreover, in any case, a simulation with the non-probabilistic model needs less elementary simulations than with Nielsen's. As a consequence, the non-probabilistic model is a better candidate to practical implementation.

Extensions of this work consist in a generalisation of these results to a universal non-approximation family.

\section{Acknowledgements}
The authors are grateful to Olivier Brunet, Marie Lalire and Mehdi Mhalla for many valuable discussions, comments and suggestions.


\end{document}